\documentclass[aps,prl,a4paper,twocolumn,showpacs,floatfix,citeautoscript,superscriptaddress,footinbib]{revtex4-1}
\usepackage{graphicx}
\usepackage{amsfonts}
\usepackage{amsmath}
\usepackage{amssymb}
\usepackage{bm}

\usepackage{bbold}
\usepackage{mathtools}
\usepackage{dcolumn}
\usepackage{hyperref}
\usepackage[usenames]{color}
\usepackage{subcaption}

\usepackage{physics}
\usepackage{units}

\usepackage{ragged2e}
\DeclareCaptionJustification{justified}{\justifying}
\newcommand{\out}[1]{{}}

\newcommand{\fref}[1]{Fig.~\ref{#1}}

\begin{document}

\title{Bulk photogalvanic current control and gap spectroscopy in 2D hexagonal materials}
\author{Anna Galler}
\email[]{anna.galler@tugraz.at}
\address{Institute of Theoretical and Computational Physics, TU Graz, Petersgasse 16, 8010 Graz, Austria}
\address{Max Planck Institute for the Structure and Dynamics of Matter, Center for Free Electron Laser Science, 22761 Hamburg, Germany}
\author{Ofer Neufeld}
\email[]{ofern@technion.ac.il}
\address{Technion Israel Institute of Technology, Faculty of Chemistry, Haifa 3200003, Israel}

\begin{abstract} 
Two-dimensional (2D) hexagonal materials have been intensively explored for multiple optoelectronic applications such as spin current generation, all-optical valleytronics, and topological electronics. In the realm of strong-field and ultrafast light-driven phenomena, it was shown that tailored laser driving such as polychromatic or few-cycle pulses can drive robust bulk photogalvanic (BPG) currents originating from the K/K' valleys. We here explore the BPG effect in 2D systems in the strong-field regime and show that monochromatic elliptical pulses also generically generate such photocurrents. The resultant photocurrents exhibit both parallel and transverse (Hall-like) components, both highly sensitive to the laser parameters, providing photocurrent control knobs. Interestingly, we show that the photocurrent amplitude has a distinct behavior vs. the driving ellipticity that can be indicative of material properties such as the gap size at K/K', which should prove useful for novel forms of BPG-based spectroscopies. We demonstrate these effects also in benchmark \textit{ab-initio} simulations in monolayer hexagonal boron-nitride. Our work establishes new paths for controlling photocurrent responses in 2D systems that can also be used for multi-dimensional spectroscopy of ultrafast material properties through photocurrent measurements.  

\end{abstract}

\maketitle

\section{Introduction}

In the last two decades, many families of two-dimensional (2D) hexagonal materials have been discovered and synthesized. This includes originally graphene~\cite{Geim2009}, and more recently transition-metal dichalcogenides (TMDs)~\cite{Manzeli2017}, hexagonal boron-nitride (hBN)~\cite{Zhang2017}, honeycomb lattices comprising elements heavier than carbon~\cite{Balendhran2015,Pumera2017}, 2D magnets~\cite{Gong2019}, and other hybrid forms~\cite{Meirzadeh2023, Evans2018, Tongay2005, PRB.92.125420, Anasori2015}. These novel structures have paved the way to a plethora of applications, from quantum information (through valleytronics~\cite{schaibley2016valleytronics, bao2022light}), topological electronics~\cite{Culcer2020, Gilbert2021}, magnetism~\cite{Gong2019}, and spintronics~\cite{Wei2016, ahn2020}. One key property of 2D hexagonal lattices is that they have a natural selective coupling to light, where the K and K' valleys in the Brillouin zone (BZ) preferentially absorb (or emit) light with a specific spin angular momentum (SAM) due to robust spin-momentum locking. This feature enables valleytronic two-level-like excitations and control through polarization tailoring, and connects topological physics with practical applications for valley and spin currents through the local Berry curvature nature of Bloch states~\cite{schaibley2016valleytronics}. 

In the realm of strong-field physics, optical selectivity was recently shown useful for manipulating the system's band structure through so-called Floquet engineering~\cite{mitra2024, jimenez2020lightwave, neufeld2023dirac, Galler2023, Neufeld2022, Neufeld2024tracking, tiwari2025robust, uzan2022observation, weitz2024lightwave}, enabling tuning of potential valley selectivity~\cite{Jimenez2021,Mrudul2021, sharma2024direct, sharma2023shaping, sharma2023giant, sharma2023valley, kim2025quantum, tyulnev2024valleytronics}, nonlinear light emission~\cite{Heide2022,zhang2024enhanced}, and lightwave-driven bulk photogalvanic (BPG) currents~\cite{Higuchi2017, neufeld2021bulk, amar2023, rana2024, schiffrin2013optical}. It is especially interesting that in each such process the underlying emitted observable (e.g. high harmonics~\cite{ghimire2019high, Yue2022}) carries valuable information about the lattice and electronic configuration that can also be used to develop novel ultrafast spectroscopies. Indeed, third-harmonic generation together with transient absorption spectra were shown useful for probing valley-polarization and valley-phonon coupling~\cite{mitra2024, Lively2024}, while high-harmonics were shown to provide information on the electronic dephasing time~\cite{kim2024, korolev2024unveiling} and argued to possibly carry topological information~\cite{Heide2022,Chen2022, Chacon2020, Baykusheva2021, schmid2021tunable, bai2021high, Bauer2018, silva2019topological}(though this remains somewhat controversial~\cite{neufeld2023topo, Rui2024}). Transport-related observables such as bulk photogalvanic currents and Hall currents were shown to provide topological information~\cite{mciver2020light, Sato2019, weitz2024lightwave, day2024nonperturbative, de2017quantized}, as well as intrinsic dynamical data connecting to electronic coherence~\cite{heide2021} and the light-matter system's symmetry~\cite{neufeld2019floquet, weitz2024lightwave, neufeld2021bulk}. 

In gapped systems with broken inversion symmetry, a bulk photogalvanic current can naturally be obtained even with monochromatic driving~\cite{Belinicher_1980,Ganichev2001}. Still, most works to date explored photocurrent generation in 2D systems with either short few-cycle pulses~\cite{Higuchi2017,schiffrin2013optical,langer2020}, or polychromatic tailored pulses~\cite{heide2021optical, neufeld2021bulk, weitz2024lightwave, sharma2025}, which provide more prominent sources of symmetry breaking that yield larger photocurrents and can also be employed in inversion-symmetric solids~\cite{neufeld2019floquet,neufeld2025}. It thus remains unclear what are the typical characteristics of the nonlinear photocurrent signals in the strong-field regime with respect to the monochromatic driving parameters (i.e. laser ellipticity and polarization axis). In optical set-ups, these parameters yield highly-sensitive harmonic signals that can be employed to probe material properties~\cite{Baykusheva2021, neufeld2023topo, yoshikawa2017high, tancogne2017ellipticity}, and if such a sensitivity exists also for BPG effect it could provide additional insight, especially through the observation of Hall-like transverse photocurrent signals that are intimately related to the Berry curvature and valley occupations~\cite{luu2018measurement, Sato2019, Yue2022}. 

Here, we numerically investigate BPG currents in model 2D hexagonal solids driven by monochromatic elliptical pulses in typical experimental laser conditions. Our calculations show that the photocurrent signal (both longitudinal and transverse components) is highly sensitive to the laser parameters, including wavelength, ellipticity, and polarization axis. In particular, the ellipticity-dependence of the signal shows a characteristic bell-like curve (similar to HHG yields in the gas phase~\cite{Horrom2012}, but maximizing at an ellipticity of ~0.5) with vanishingly-small photocurrents for linear and circular driving (due to symmetry), and where the signal is modulated in width and appearance of multiple peaks by tuning the laser parameters or material properties. Remarkably, we show that the photocurrent behavior with driving ellipticity has a distinct structure that changes sign multiple times as the system inversion symmetry is broken (i.e. transitioning from gapless to gaped solid), with linear scaling for small gap size. We further validate these results with \textit{ab-initio} simulations in hBN. Our work therefore establishes the fundamental transport response to elliptical driving in the strong-field regime, and paves way to novel forms of ultrafast spectroscopy of dynamical material properties based on photocurrent measurements (e.g. the gap, which can be relevant and difficult to measure in Floquet systems~\cite{rudner2020band, broers2021observing, merboldt2024observation, aeschlimann2021}). 

\begin{figure}[t!]
\includegraphics[width=\columnwidth]{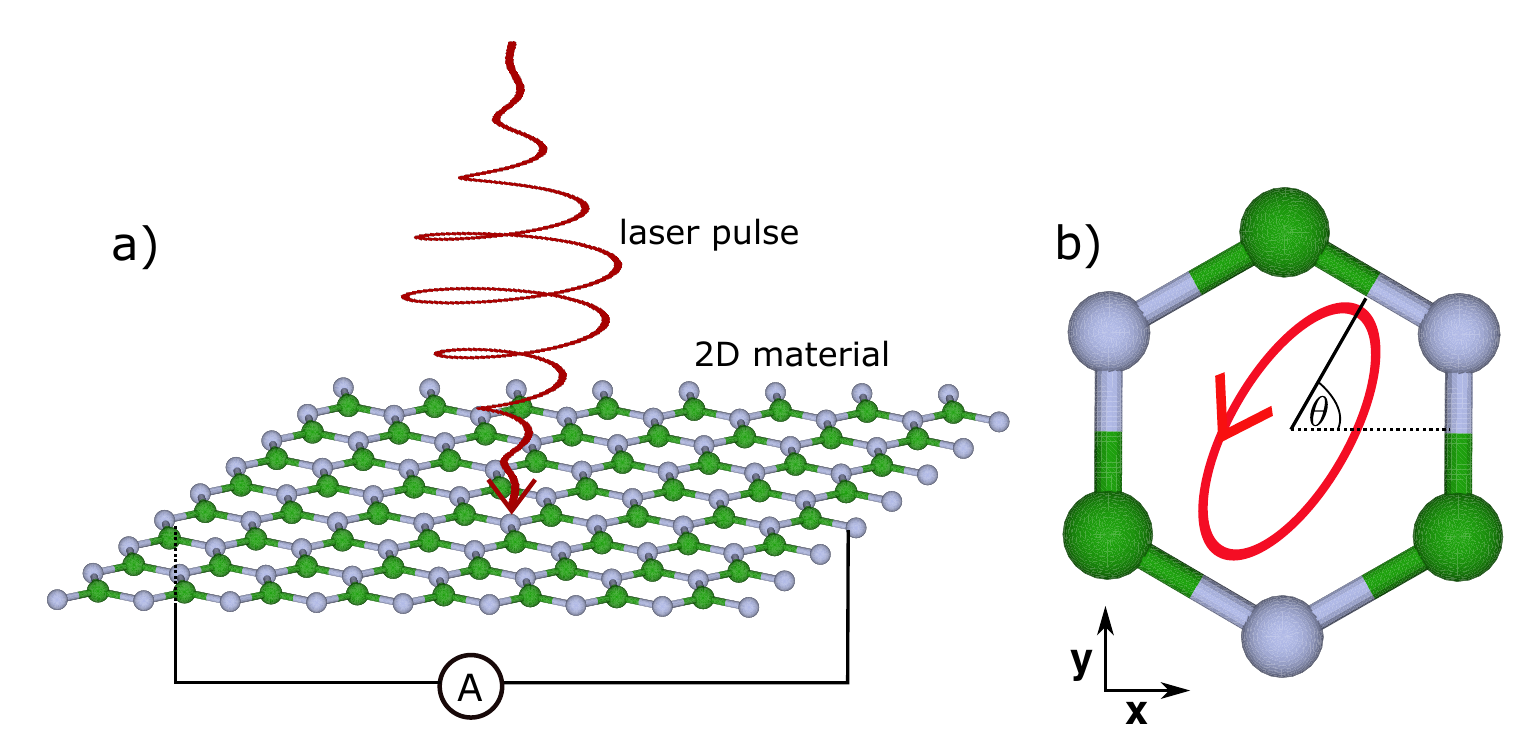}
 \caption{Photocurrent generation in a hexagonal 2D material. (a) Elliptically polarized laser pulse incident on a monolayer of hexagonal boron nitride (hBN). (b) Orientation $\theta$ of the major elliptical axis with respect to the crystal lattice. $\theta=0$ corresponds to the zigzag direction of the lattice.} 
\label{fig:sketch}
\end{figure}

\section{Methods}
We investigate photocurrents in a generic two-dimensional model system with valley degrees of freedom, and in a realistic 2D material---a monolayer of hexagonal boron nitride (hBN). Below we describe the methodology of each approach.

\subsection{Model calculations}
 We employ a real-space model of a honeycomb lattice with A/B sublattice sites and periodic boundary conditions. Each site is represented by a local Gaussian potential given in atomic units by
\begin{equation} 
\label{eq:gauss}
V_{A,B}(\textbf{r})=-v_{0,A,B} e^{-\textbf{r}^2/\sigma_{A,B}^2}, 
\end{equation}
where $\textbf{r}$ is the electronic coordinate in 2D. We choose $v_{0,A}=40$ eV and $\sigma_{A}=1.5$ Bohr, while the Gaussian potential on lattice site B can be varied from $v_{0,B}=v_{0,A}$ to $v_{0,B}=1.08v_{0,A}$, to interpolate between a gapless model and a model with up to $\unit[2]{eV}$ band gap at the $K/K'$ points. The lattice vectors employed are of length $2\pi$ Bohr of the form $\mathbf{a_1}=2\pi\hat{\mathbf{x}}$, $\mathbf{a_2}=2\pi(-\hat{\mathbf{x}}/2+\sqrt(3)\hat{\mathbf{y}}/2)$.
Each site contributes one electron per unit cell, resulting in two electrons per cell that occupy the first valence band. Spin-orbit coupling and electron-electron interactions are neglected. The field-free Hamiltonian of the system, in atomic units and real-space representation, is given by
\begin{align} 
h_{0}(\textbf{r}) &= \sum_{n,m}V_{A}(\mathbf{r}-n\mathbf{a_1}-m\mathbf{a_2})\nonumber \\ &+\sum_{n,m}V_{B}(\mathbf{r}-n\mathbf{a_1}-m\mathbf{a_2}-\frac{\mathbf{a_1}}{3}+\frac{\mathbf{a_2}}{3}) -\frac{1}{2}\nabla^2 \nonumber \\
&=V(\mathbf{r},\mathbf{a_1},\mathbf{a_2})-\frac{1}{2}\nabla^2 ,
\end{align}
where $n$ and $m$ are integers. We apply periodic boundary conditions along the lattice vectors and diagonalize the Hamiltonian to obtain the ground-state Bloch states, using a real-space grid spacing of $\unit[0.28]{Bohr}$ and a $k$-space discretization with a $\Gamma$-centered $100\times100$ $k$-grid. 

For equivalent sublattice sites with $v_{0,B}=v_{0,A}$ and $\sigma_{B}=\sigma_{A}$,  this setup produces a gapless band structure with Dirac cones at the $K/K'$ points resembling the electronic band structure of graphene. For differing sublattices A/B, we obtain a honeycomb lattice with broken inversion symmetry, featuring a direct optical gap at the $K/K'$ points. For our chosen parameter set, the direct optical gap ranges from $\unit[0-2]{eV}$. To investigate laser-induced electronic dynamics, we simulate the interaction of this system with an intense elliptically polarized laser pulse (up to $\sim$0.3 TW cm$^{-2}$) with a non-resonant carrier frequency well below the band gap. Our numerical approach involves solving the time-dependent Schrödinger equation (TDSE) under the dipole approximation while assuming the independent particle approximation (neglecting electron-electron interactions).

\begin{figure}[t!]
\includegraphics[width=\columnwidth]{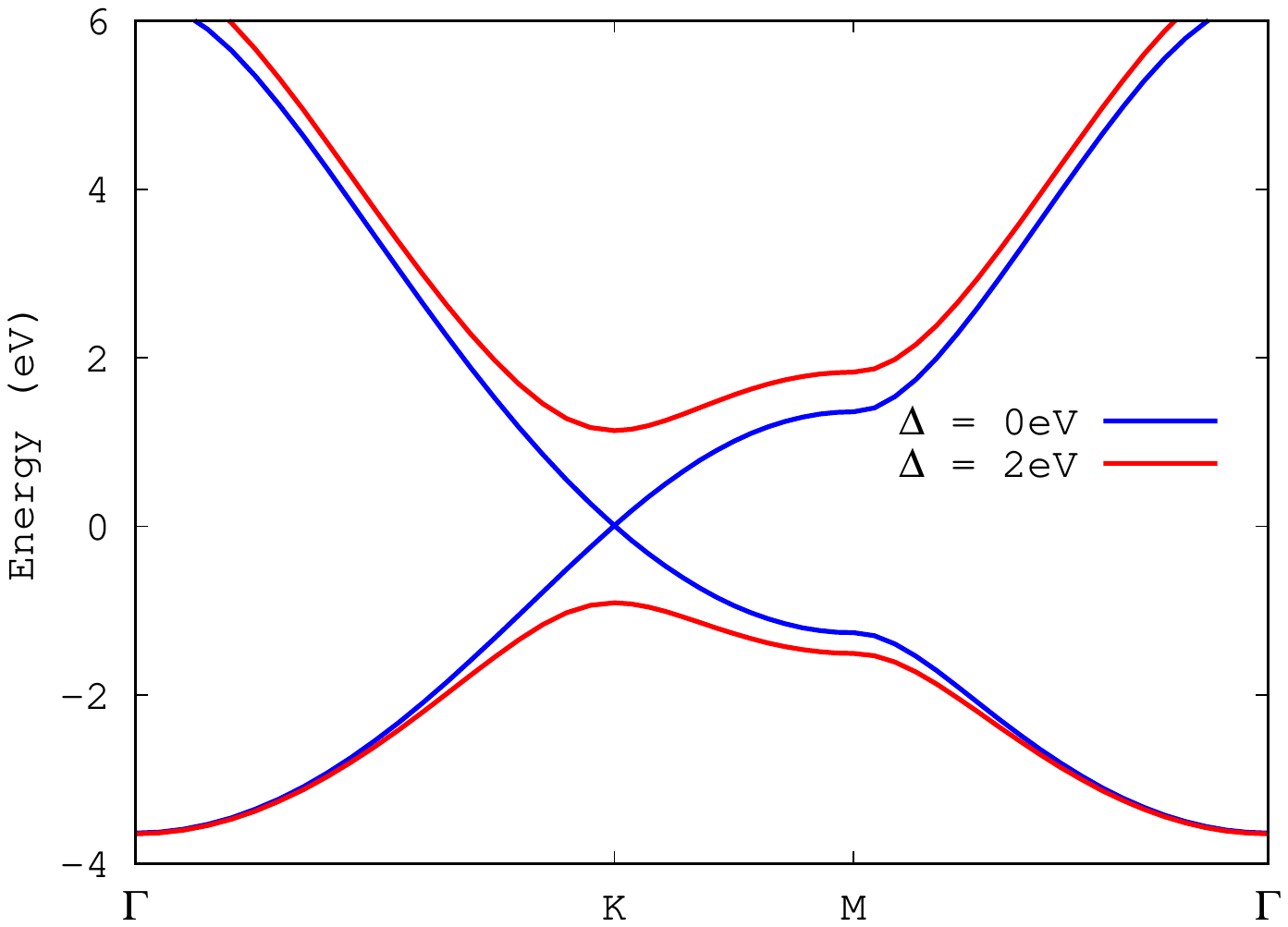}
 \caption{Band structure of the employed honeycomb model. For equivalent sublattice sites with $v_{0,B}=v_{0,A}$, one obtains a gapless electronic band structure (in blue) with a Dirac cone at $K$. By breaking the inversion symmetry of the lattice through differing $A/B$ sublattice sites with $v_{0,B}=1.08v_{0,A}$, a gap of $\unit[2]{eV}$ is opened at $K$ (in red).} 
\label{fig:bands}
\end{figure}

The TDSE is solved numerically by propagating all initially-occupied Bloch states on the real-space grid representation:
\begin{align} 
i\partial_t\ket{\psi_{k}(t)} &=\left[V(\mathbf{r},\mathbf{a_1},\mathbf{a_2}) +\frac{1}{2}(-i\nabla-\frac{1}{c}\mathbf{A}(t))^2\right]\ket{\psi_{k}(t)} ,
\end{align}
where we employ the velocity gauge for the laser-matter interaction, and $\mathbf{A}(t)$ is the vector potential of the laser electric field such that $-\partial_t\mathbf{A}(t)=c\mathbf{E}(t)$, with $c$ being the speed of light. The ground state of the model is taken as the initial state and propagation is performed with a Lanczos expansion method with a time-step of $\unit[0.2]{a.u.}$ The laser vector potential employed is taken as
\begin{equation}
\label{eq:a}
\mathbf{A}(t)=f(t)\frac{cE_{0}}{\omega \sqrt{1+\epsilon^2}}\hat{R(\theta)}\cdot(\cos(\omega t)\hat{\mathbf{x}}+\epsilon \sin(\omega t)\hat{\mathbf{y}}) ,
\end{equation}
where $E_0$ is the electric field amplitude, $\omega$ is the driving frequency,  $\epsilon$ is the field's ellipticity, and $\hat{R(\theta)}$ is a rotation matrix in the $xy$-plane operating on the polarization vector. Overall, eq.~\eqref{eq:a} describes an elliptically-polarized laser pulse with an elliptical major axis oriented $\theta$ degrees above the $x$-axis. $f(t)$ in Eq.\eqref{eq:a} is a normalized temporal envelope function taken as a super-sine~\cite{neufeld2019} form
\begin{equation}
f(t) = \sin{\left(\pi \frac{t}{T_p}\right)}^{\left(\frac{|\pi(\frac{t}{T_p}-\frac{1}{2})|}{\sigma}\right)},
\label{eq:env}
\end{equation}
where $\sigma=0.75$, $T_p$ is the duration of the laser pulse which was chosen as $T_p=10T$, where T is a single cycle of the fundamental carrier frequency (the overall full-width-half-max of the pulse is 5T).

The time-dependent current expectation value is given by
\begin{equation}
\mathbf{J}(t) = \frac{1}{V} \int_V \mathbf{j}(\mathbf{r},t)d^3r,
\end{equation}
where $V$ denotes the volume of the primitive unit cell and $\mathbf{j}(\mathbf{r},t)$  represents the microscopic time-dependent current density, defined as
\begin{equation}
\label{eq:curr}
\mathbf{j}(\mathbf{r},t) = \frac{1}{2} \sum_{k} \left[ \psi_{k}^*(\mathbf{r},t) \left( -i\nabla - \frac{\mathbf{A}(t)}{c}  \right) \psi_{k}(\mathbf{r},t) + c.c. \right].
\end{equation}
The photocurrent $\mathbf{J}(t)$ is averaged over multiple laser cycles after the pulse ended: six cycles  for \unit[800]{nm} pulses and two cycles for \unit[3000]{nm} pulses.

\begin{figure*}[t!]
\begin{minipage}{17. cm}
\includegraphics[width=\columnwidth]{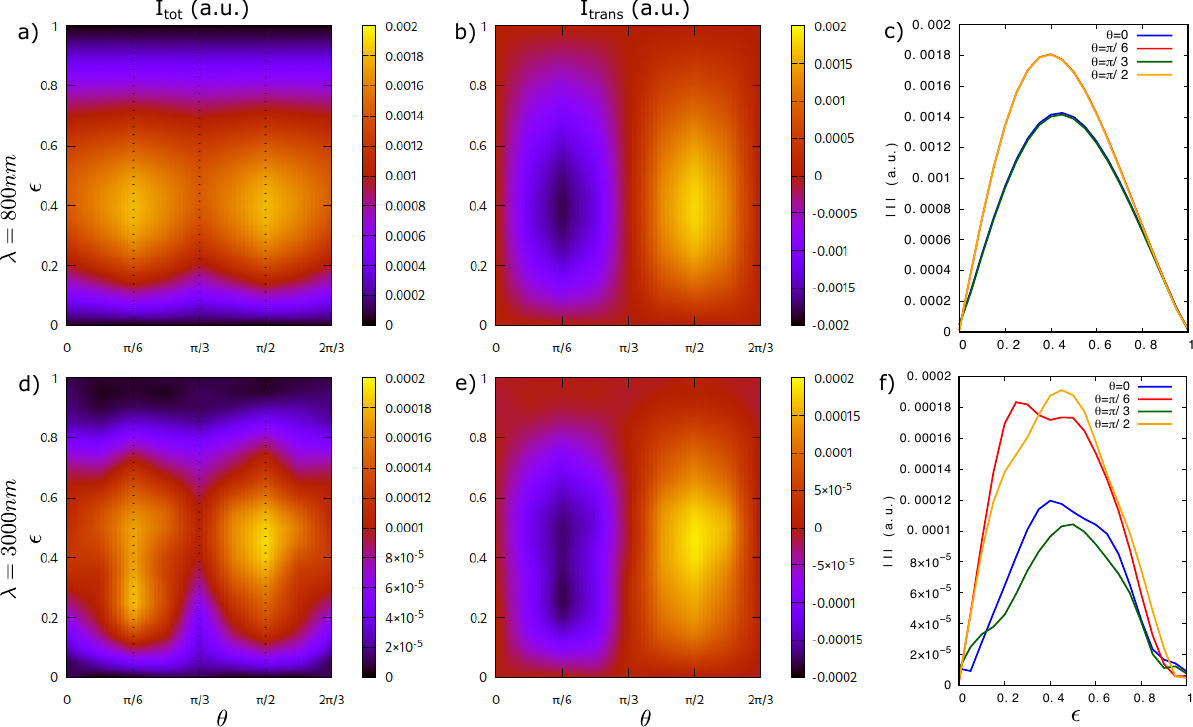}
\caption{Photocurrent in a 2D hexagonal lattice with broken inversion symmetry and a band gap of $\unit[2]{eV}$. (a) Amplitude of the photocurrent  induced by an elliptically polarized laser pulse with ellipticity $\epsilon$, orientation $\theta$  of the main elliptical axis and wavelength $\lambda=\unit[800]{nm}$. (b) A transverse current is induced with maxima for $\theta=\pi/6$ and $\theta=\pi/2$. (c) Line-cuts of the total current for $\theta=0,\pi/6,\pi/3,\pi/2$. (d-f) Corresponding results for a driving laser pulse with $\lambda=\unit[3000]{nm}$.   } 
\label{fig:current_multi}
\end{minipage}
\end{figure*}

\subsection{TDDFT calculations}
Time-dependent density functional theory (TDDFT) calculations are performed using the real-space grid-based code \verb=Octopus=~\cite{Octopus2015,Octopus2020}. The Kohn-Sham (KS) equations are discretized on a 3D Cartesian grid within the primitive unit cell of monolayer hBN using the experimental lattice parameter a=\unit[2.52]{\AA}. A vacuum spacing of \unit[40]{Bohr} is included above and below the monolayer to prevent spurious interactions.
Calculations are performed using the local density approximation (LDA), neglecting spin degrees of freedom and spin-orbit coupling. The frozen-core approximation is applied, with inner core states treated using norm-conserving pseudopotentials~\cite{hartwigsen1998relativistic}. The KS equations are solved self-consistently with an energy convergence threshold of $<\unit[10^{-7}]{Hartree}$, and the real-space grid spacing is converged to \unit[0.37]{Bohr}. A $\Gamma$-centred $k$-grid with 72$\times$72 $k$-points is used for Brillouin zone sampling.

In the TDDFT calculations, the time-dependent Kohn-Sham (KS) equations are solved within the adiabatic approximation and in the velocity gauge. In atomic units, the KS equations are given by
\begin{equation}
i\partial_t\ket{\psi_{nk}^{KS}(t)} = \left[\frac{1}{2}\left(-i\nabla-\frac{\mathbf{A}(t)}{c} \right)^2 + v(\textbf{r},t) \right]\ket{\psi_{nk}^{KS}(t)} ,
\end{equation}
where $\ket{\psi_{nk}^{KS}(t)}$ are the KS single-particle wave functions for band  $n$ and $k$-point $k$, and $\mathbf{A}(t)$ is the vector potential of the laser field as in Eq.~\eqref{eq:a}.
$v(\textbf{r},t)$ is the time-dependent KS potential given by 
\begin{equation}
\label{eq:ks_pot}
v(\textbf{r},t) = -\sum_I\frac{Z_I}{|\mathbf{R}_I-\mathbf{r}|}+\int d^3r'\frac{n(\mathbf{r'},t)}{|\mathbf{r}-\mathbf{r'}|}+v_{XC}[n(\mathbf{r},t)] ,
\end{equation}
where $Z_I$ and $\mathbf{R}_I$  denote the charge and position of the $I$th nucleus, respectively. $v_{XC}$ represents the exchange-correlation potential, which is a functional of the time-dependent electron density  $n(\mathbf{r},t)=\sum_{nk}|\ket{\psi_{nk}^{KS}(t)}|^2$. In practice, the Coulomb interaction with the nuclei (first term in Eq.~\eqref{eq:ks_pot}) is replaced by a non-local pseudopotential that also accounts for core electron contributions.
The KS wave functions are propagated with a similar approach as in the model system, and with the same approach applied to obtain the photocurrent signal.

\section{Results and discussion}

\subsection{Ellipticity-dependent photocurrent}
Our examined setup is illustrated in \fref{fig:sketch}a --- an elliptically polarized laser pulse is incident on a 2D hexagonal material with broken inversion symmetry and polarized in the monolayer plane. The laser pulse generates a bulk photogalvanic current that can be detected macroscopically by transport measurements. \fref{fig:sketch}b further illustrates the orientation of the light polarization, with $\theta$ the angle between the major axis of the laser’s elliptical polarization and the zigzag lattice direction.

We begin our investigation focusing on the hexagonal lattice model with broken inversion symmetry and a direct band gap of $\unit[2]{eV}$ at $K/K'$. The electronic band structure of this model is shown in \fref{fig:bands} (in red).
In \fref{fig:current_multi}, we present the induced BPG current for an elliptically polarized laser pulse with wavelengths of $\lambda=\unit[800]{nm}$ (top row) and $\lambda=\unit[3000]{nm}$ (bottom row). The color maps in subfigures (a–b) and (d–e) illustrate the photocurrent as a function of the laser ellipticity $\epsilon$ and the orientation $\theta$ of the major elliptical axis. Notably, no photocurrent is generated for linearly ($\epsilon=0$) or circularly ($\epsilon=1$) polarized light. Instead, the BPG current reaches its maximum absolute value for ellipticities in the range $\epsilon=0.3-0.6$, as evident from the color map in \fref{fig:current_multi}a and the corresponding line-outs in \fref{fig:current_multi}c, which exhibit distinct bell-shaped characteristic curves.

\begin{figure}[t!]
\includegraphics[width=\columnwidth]{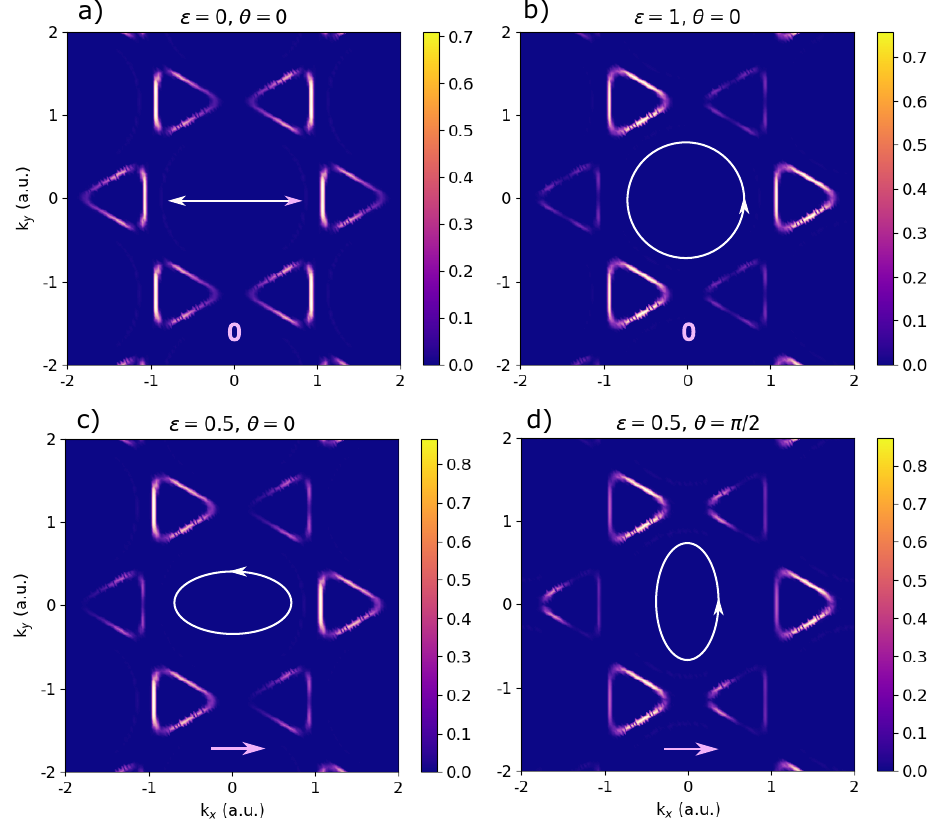}
 \caption{Momentum-resolved charge distribution in the conduction band at the end of the laser pulse. Shown are results for (a) linear, (b) circular and (c-d) elliptical polarization, $\unit[800]{nm}$ wavelength and $\unit[0.3]{TW/cm^2}$ laser intensity.  There is no resulting photocurrent for (a) linear and (b) circular driving due to a K/K' mirror plane and a threefold warping symmetry in the K/K' valley, respectively. An elliptical pump pulse with $\epsilon=0.5$, $\theta=0,\pi/2$ yields a photocurrent in zigzag direction, which, for elliptical driving with $\theta=\pi/2$ is fully transverse.} 
\label{fig:pattern}
\end{figure}

Interestingly, the absolute maximum of the BPG current occurs when the major elliptical axis is oriented at $\theta=\pi/6$ and $\theta=\pi/2$ relative to the zigzag direction of the crystal lattice. A decomposition of the current into its longitudinal and transverse components (\fref{fig:current_multi}b) reveals that at these orientations, the BPG current is entirely transverse and flows in opposite directions for $\theta=\pi/6$ and $\theta=\pi/2$. Specifically, when the major elliptical axis is aligned along the armchair direction ($\theta=\pi/2$) of the hexagonal lattice, the photocurrent flows perpendicularly, i.e., along the zigzag direction. Conversely, when the major elliptical axis is aligned with the zigzag direction, the current remains entirely longitudinal along the same direction. This is a result of fundamental mirror symmetries in the hexagonal lattice and will be discussed below.

For longer-wavelength driving at $\lambda=\unit[3000]{nm}$, the overall trends remain similar (see \fref{fig:current_multi}d–f). However, additional features such as double-peak structures emerge, likely due to band structure effects beyond K/K' being probed by the longer laser drive (since longer wavelengths correspond to an overall higher amplitude vector potential that drives electrons further across the Brillouin zone).

The observed behavior of the BPG current can be understood in terms of the symmetries of both the hexagonal lattice and the laser polarization. In \fref{fig:pattern}, we illustrate the $k$-resolved electron distribution in the conduction band (CB) at the end of the laser pulse for (a) linear, (b) circular, and (c–d) elliptical driving. The CB occupations are computed by projecting the time-dependent Bloch state onto the ground-state CB wavefunction, $g_{CB}(\mathbf{k})=|\bra{\psi_{CB,\mathbf{k}}(t=0)}\ket{\psi_{\mathbf{k}}(t_{end})}|^2$.
The resulting ring-shaped charge distributions follow multi-photon resonant contours of the Floquet light-dressed electronic states, as discussed in detail in Ref.~\cite{Galler2023}. Here, we focus on the symmetry properties of these laser-induced charge patterns, which directly determine the emergence of BPG currents. In general, an imbalance in the $k$-space charge distribution in the CB induces a photocurrent.

For a linearly polarized laser (\fref{fig:pattern}a), the $k$-resolved CB occupation exhibits a mirror symmetry across both $k_x=0$ and $k_y=0$, leading to a vanishing net BPG current. For circularly polarized driving (\fref{fig:pattern}b), the CB charge occupation displays a threefold warping symmetry within each $K$ and $K'$ valley, which effectively suppresses any BPG currents as well. Instead, the well-known valley polarization effect is obtained, with the $K$ valley appearing bright and the $K'$ valley dark.
In contrast, elliptically polarized laser pulses induce an asymmetric CB charge distribution that results in a net photocurrent by virtue of their broken symmetry. For example, when the major elliptical axis is aligned along the zigzag direction ($\theta=0$), the CB charge pattern retains only a mirror plane at $k_y=0$ (\fref{fig:pattern}c), generating a photocurrent along the zigzag ($k_x$) direction. When the major elliptical axis is rotated to $\theta=\pi/2$, as shown in \fref{fig:pattern}d, the CB charge distribution still exhibits mirror symmetry across $k_y=0$, and the resulting photocurrent now flows along $k_x$, perpendicular to the ellipse's major axis. This analysis is in agreement with the observed BPG currents in \fref{fig:current_multi}.

More generally, our results demonstrate that the photocurrent in a 2D hexagonal material with broken inversion symmetry can be effectively controlled by tuning the ellipticity $\epsilon$ and in-plane orientation $\theta$ of an elliptically polarized laser, which is facilitated by a fine-tuned control of the excited electron occupation patterns in $k$-space.

\subsection{Gap spectroscopy}
The ellipticity-dependent BPG currents discussed so far were computed for a honeycomb model with a $\Delta=\unit[2]{eV}$ optical gap at $K$ and $K'$ (roughly modeling TMDs). To explore the impact of the gap size on the photocurrent, we systematically varied the band gap by adjusting the Gaussian potentials on sublattice sites $A$ and $B$ (see Eq.\eqref{eq:gauss}). When the sublattice potentials are identical, $v_{0,A}=v_{0,B}$, the system is gapless, exhibiting a Dirac cone at $K$ and resembling the band structure of graphene (see \fref{fig:bands}). In contrast, when $v_{0,A}\neq v_{0,B}$, inversion symmetry is broken, opening a direct gap at $K$ and $K'$ in the electronic band structure.

\begin{figure*}[t!]
\begin{minipage}{17. cm}
\includegraphics[width=\columnwidth]{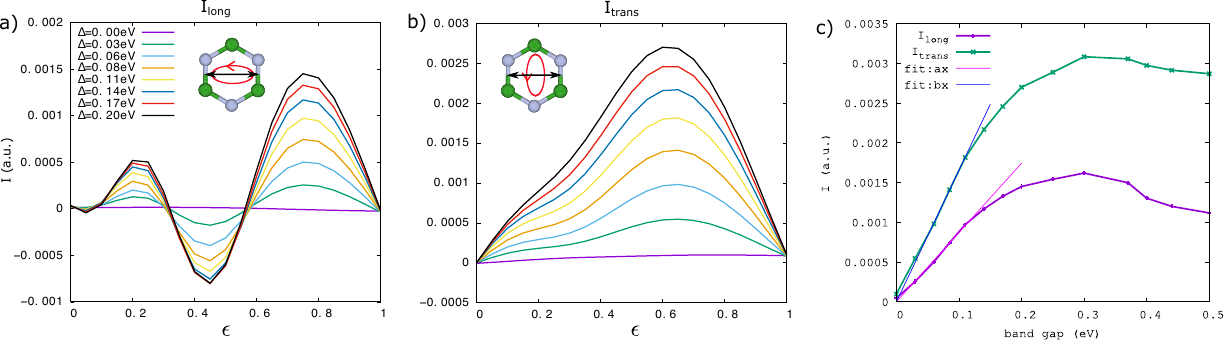}
\caption{Dependence of the photocurrent on the size of the band gap. (a) Longitudinal BPG current, induced by elliptical driving along the zigzag direction (inset), for band gaps in the range $\Delta=\unit[0-0.2]{eV}$. The longitudinal current changes direction multiple times depending on the ellipticity $\epsilon$. (b) The transverse current, induced by elliptical driving along the armchair direction (inset), increases monotonously with the gap size. (c) Scaling of the maximum value of the longitudinal and transverse photocurrent. For small band gaps up to $\unit[0.12]{eV}$, the photocurrent depends linearly on the gap size and can be well approximated by a linear fit with $a=0.0087$ (purple line) and $b=0.0165$ (blue line), respectively. } 
\label{fig:model_osc}
\end{minipage}
\end{figure*}

Figure~\ref{fig:model_osc} presents the computed BPG currents for band gaps in the range of $\Delta=\unit[0-0.2]{eV}$. As expected, no photocurrent is observed in the inversion-symmetric, gapless case. This is a consequence of inversion symmetry being respected in both the materials system, and any elliptical laser pulse precluding second-order nonlinear responses such as shift currents~\cite{neufeld2019floquet}. For small gaps up to \unit[0.2]{eV}, the current gradually increases, as shown in \fref{fig:model_osc}, presenting a telltale sign for symmetry breaking. Interestingly, the longitudinal current (\fref{fig:model_osc}a) induced by elliptical driving along the zigzag direction (inset of \fref{fig:model_osc}a) reverses direction multiple times depending on the driving ellipticity $\epsilon$. In contrast, the transverse current (\fref{fig:model_osc}b) exhibits a unidirectional increase. The multiple sign changes vs. driving ellipticity in Fig. 5(a) is highly counter-intuitive in the sense that light's helicity remains unchanged throughout. In other words, semi-classical dynamics of band electrons are not expected to cause alternating photocurrent sign changes. Still, the sign changes reflect the inherent symmetry breaking in the electronic bands, potentially connecting to multiple interfering pathways for photocurrent generation that open up as the Dirac cone gaps out. As the gap further increases, these signals vanish and are replaced by a clear preferred directionality for photocurrents for any ellipticity value.

The scaling of the BPG current is further analyzed in \fref{fig:model_osc}c. For small band gaps up to \unit[0.12]{eV}, both the longitudinal and transverse components of the current exhibit a linear increase with gap size. As demonstrated in \fref{fig:model_osc}c, this trend is well captured by a linear least-squares fit, indicating that a linearly increasing BPG current serves as a clear signature of band-gap opening in 2D hexagonal materials, such as graphene subjected to Floquet engineering~\cite{merboldt2024observation,weitz2024lightwave}. However, for larger gaps exceeding \unit[0.12]{eV}, the linear relationship breaks down, and accurately describing the current-gap dependence requires polynomial fits of at least third order.

\begin{figure}[t!]
\includegraphics[width=\columnwidth]{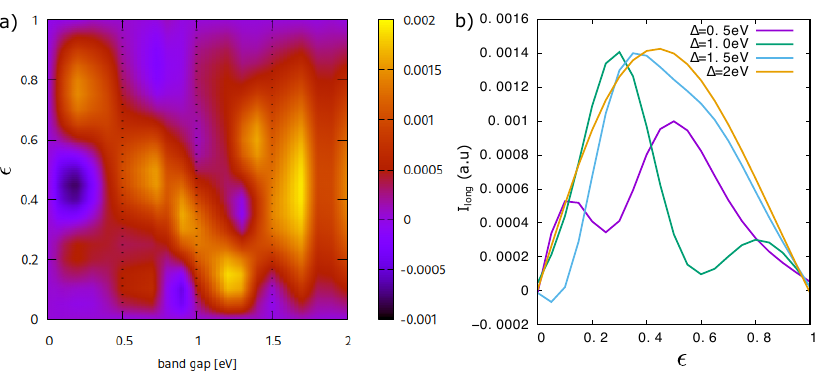}
\caption{(a) Longitudinal BPG current $I_{long}$ depending on ellipticity $\epsilon$ and gap size $\Delta$ (with the laser's major elliptical axis along the zigzag direction). The behaviour of the BPG current can essentially be divided into three regions: (i) $\Delta=\unit[0-0.3]{eV}$, (ii) $\Delta=\unit[0.3-1.5]{eV}$, and (iii) $\Delta=\unit[1.5-2]{eV}$. (c) While the ellipticity-dependent line-outs in region (ii) show a two-peak structure, a single-peak structure with a maximum around $\epsilon=0.5$ develops above $\Delta=\unit[1.5]{eV}$.  } 
\label{fig:model_gap}
\end{figure}

In \fref{fig:model_gap}, we further examine the evolution of the BPG current for larger band gaps above \unit[0.5]{eV}. \fref{fig:model_gap}a presents the longitudinal BPG current as a function of ellipticity $\epsilon$ and gap size $\Delta$. The photocurrent behavior can be categorized into three distinct regimes. The small-gap regime ($\Delta=\unit[0-0.3]{eV}$) as has already been discussed above. In the intermediate-gap range ($\Delta=\unit[0.3-1.5]{eV}$), the photocurrent develops a double-peak structure as a function of $\epsilon$. Above $\Delta=\unit[1.5]{eV}$, this double-peak pattern transitions into a single-peak, bell-shaped curve, a trend clearly visible in the line-outs shown in \fref{fig:model_gap}b. The physical origin for the formation of the double peak currently remains unclear to us and should motivate future research, though we suspect it arises as a result of interference between several multi-photon channels for photocurrent generation that open up as the gap opens (while in small gaps a single channel near to K/K' points should be dominant). Thus, the double-peak feature could potentially be used as a fingerprint sign for quantum interference spectroscopy through BPG measurements. 
  
\subsection{Monolayer hBN}
Next, we extend our analysis also towards a realistic 2D material. We perform \textit{ab-initio} calculations for a monolayer of hBN irradiated by an intense laser pulse with a wavelength of $\unit[800]{nm}$. Note that the methodology and conditions here are similar to the one employed for the model, but incorporates multiple optically active valence electrons that interact both with each other and with the driving laser field. For comparison, we further perform model calculations for the aforementioned hexagonal model with a gap of \unit[4.2]{eV}, which corresponds to the band gap in hBN at the LDA level.

\begin{figure}[t!]
\includegraphics[width=\columnwidth]{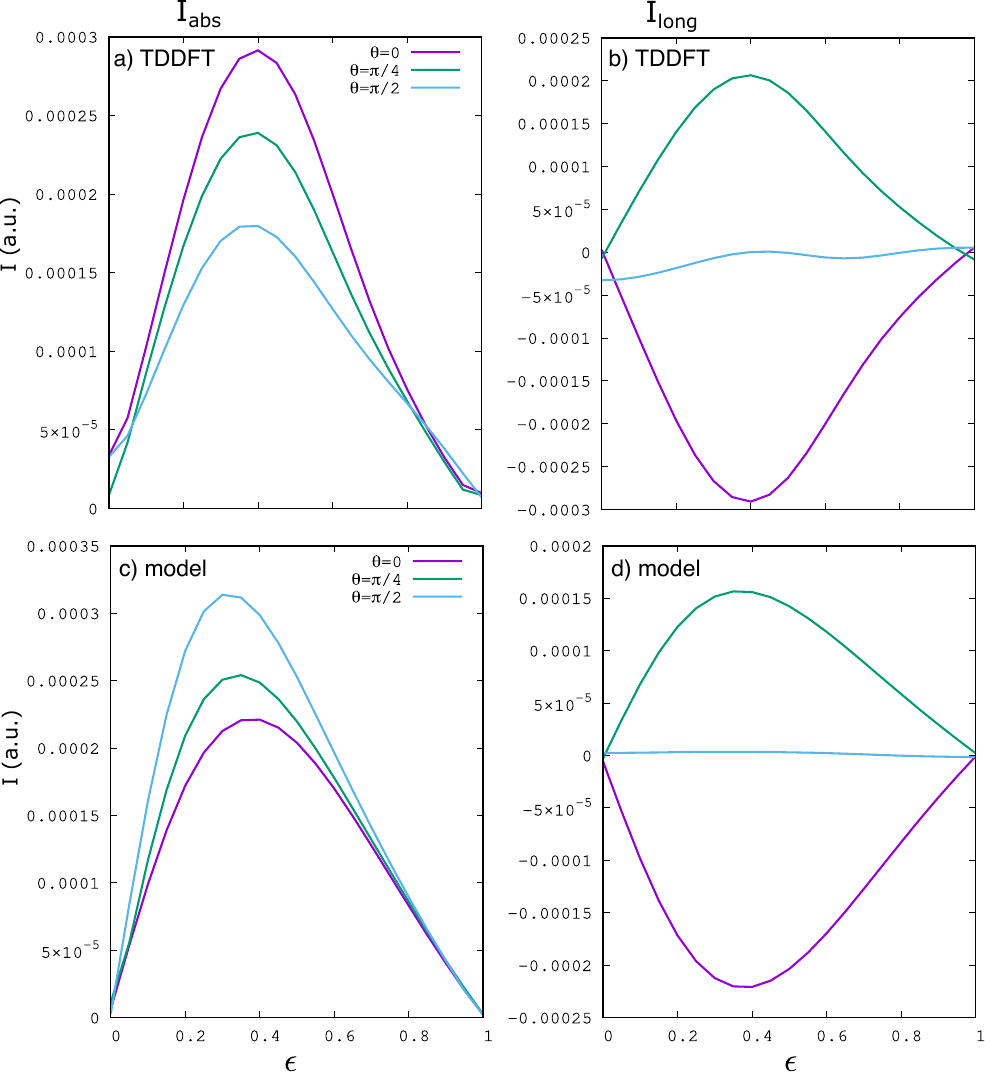}
\caption{BPG current in a monolayer of hexagonal boron nitride (hBN). (a) In TDDFT, the absolute value of the photocurrent exhibits bell-shaped maxima at ellipticities around $\epsilon=0.4$. (b) The longitudinal component of the photocurrent reverses direction when the major axis of the light's ellipticity rotates from $\theta=0$ (violet) to $\theta=\pi/4$ (green). For $\theta=\pi/2$ (blue), the longitudinal current vanishes. (c-d) The corresponding results for a hexagonal model with a $\Delta=\unit[4.2]{eV}$ band gap agree well with the TDDFT simulations.} 
\label{fig:hbn}
\end{figure}

\fref{fig:hbn} presents computed BPG currents from monolayer hBN obtained for various driving conditions using (a–b) TDDFT and, for comparison, the computationally less demanding (c–d) model calculations. As shown in \fref{fig:hbn}a, the TDDFT results exhibit pronounced bell-shaped maxima around $\epsilon = 0.4$, in close agreement with the model predictions in \fref{fig:hbn}c. The primary difference lies in the orientation that yields the maximum current: in the TDDFT simulations, the maximum occurs for elliptical driving along $\theta = 0$, corresponding to the zigzag direction of the lattice, while in the model, the maximum is found for driving along the armchair direction ($\theta = \pi/2$). We believe that this discrepancy arises from the multi-band nature of the \textit{ab-initio} simulations compared to the model system, as well as discrepancies in the band structure away from the K/K' valleys. Both of these effects can slightly alter nonlinear responses in intense laser driving conditions. Notably, the longitudinal component of the photocurrent, shown in \fref{fig:hbn}b, reverses its direction when the major axis of the ellipse is rotated from $\theta = 0$ to $\theta = \pi/4$. For $\theta = \pi/2$, where the polarization aligns with the armchair direction of the hexagonal lattice, the longitudinal current component vanishes, which is consistent with the model results and supports other data for enabling nonlinear photocurrent spectroscopy of material systems. Most importantly, the excellent agreement between the model and TDDFT simulations supports the validity of the model results presented in the previous sections for realistic 2D hexagonal materials such as hBN and monolayer transition-metal dichalcogenides, e.g. for gap spectroscopies and in inducing unique ellipticity-dependent features in the photocurrent response.

\section{Conclusions}
To summarize, we numerically explored nonlinear photogalvanic currents driven in hexagonal 2D systems by elliptically-polarized intense laser pulses. From model and \textit{ab-initio} simulations performed at various interaction and material regimes, we established that light-driven photocurrents are highly sensitive to the parameters of the laser, as well as material properties. In particular, we showed that the induced current amplitude (both longitudinal and transverse Hall) has a characteristic bell-like curve with respect to the driving laser ellipticity, which is further modulated by tuning the main elliptical axis compared with high-symmetry planes. This dependence complexifies with longer laser wavelengths, yielding multiple peak structures. Strikingly, as the system transitions from centrosymmetric to non-centrosymmetric (opening a bulk gap with non-vanishing Berry curvature), the photocurrent signal vs. ellipticity shows emergence of negative and positive peaks with an amplitude that scales linearly with the laser power. This behavior is not expected from semi-classical band dynamics, and likely reflects multiple interfering pathways for photocurrent generation. The size of the photocurrent scales linearly with the gap size, providing an indirect probe of the band structure.  

These unique signatures in the photocurrent behavior could be useful for developing novel ultrafast spectroscopies of topological systems, e.g. Floquet topological insulators or topological surfaces states in the presence of defects and scatterers. Thus, our work promotes the application of light-driven photocurrents as a novel probe of ultrafast dynamics and material structure, placing it as a complementary approach to all-optical schemes such as HHG-spectroscopy and transient absorption.

\section{Acknowledgments}
The authors gratefully acknowledge the scientific support provided by Prof.~\'Angel Rubio. This work was financially supported by the Austrian Science Fund (FWF) grant 10.55776/V988.

%


\end{document}